\documentclass[aps,prl,reprint,showpacs,showkeys,groupedaddress]{revtex4-1}
\usepackage{amsmath, amssymb}
\usepackage{amsmath}
\usepackage{graphicx}
\usepackage{graphics}
\usepackage{subfigure}
\usepackage{color}
\usepackage[colorlinks=true,allcolors=blue]{hyperref}
\usepackage{float}

\newcommand*{\rom}[1]{}

\begin{document}

%\preprint{}

\title{Flow cross-overs under surface fluctuations in cylindrical nano-channel}
\author{Aakash Anand}\thanks{aakash.a@students.iiserpune.ac.in}
%\thanks{}
\author{A. Bhattacharyay}\thanks{a.bhattacharyay@iiserpune.ac.in}
\affiliation{Department of Physics, Indian Institute of Science Education and Research, Pune, Maharashtra 411008, India.}

\date{\today}

\begin{abstract}
We analyse surface-fluctuations-driven fluid flow through nano-channels to investigate the interplay between boundary layer flow structures and the bulk flow of fluid under a pressure-head. Surface fluctuations of a wide range of frequencies (up to several thousands of Hertz) in a nano-channel keep the flow in the low Reynolds number regime. Using this advantage of low Reynolds number flow, we develop a perturbation analysis of the fluid flow that clearly distinguishes the bulk flow under a pressure head around the axis of a nano-tube from its surface flow structure induced by fluctuations. In terms of particle transport under such flow conditions, there exists the opportunity to drag particles near the periphery of the nano-tube in a direction opposite to the bulk flow near the axis. This can potentially find applications in the separation, trapping, and filtration of particles under surface-driven flow through nano-tubes under widely varying conditions.

\end{abstract}
\pacs{}
%\keywords{}

\maketitle

%\section{Introduction}
\section{INTRODUCTION}
In the realm of microfluidics and nanofluidics \cite{whitesides2006origins,kirby2010micro, kirby2011micro, squires2005microfluidics, tabeling2023introduction} the manipulation and control of fluid flow at the smallest scales hold potential for various applications, ranging from lab-on-a-chip\cite{stone2004engineering, gupta2016lab, stein2006pressure} devices to drug delivery\cite{akgonullu2021microfluidic} systems and beyond. This paper delves into the dynamics of surface-fluctuations-driven fluid flow\cite{marbach2018transport} through nano-channels, presenting a detailed analysis that distinguishes boundary layer flow structures from the bulk flow of fluid under a pressure-head. The investigation explores surface fluctuations spanning a wide range of frequencies, effectively maintaining the flow in the low Reynolds number regime.

Harnessing the advantages of low Reynolds number flow\cite{happel1983low}, the paper employs a perturbation analysis to differentiate between bulk flow around the axis of a nano-tube and the surface flow structure induced by fluctuations. We present the quantification of the surface-driven flow in terms of fluid and driving parameters(density, frequency, amplitude and wavelength). Also, we present the possibility of tuning the longitudinal and transverse modes of transport by changing the wavelength of the driving. This paper presents the methodology, findings, and implications of the study in detail.

The unique characteristics of this flow regime create opportunities for particle transport near the periphery of the nano-tube, counter to the bulk flow near the axis. This intriguing phenomenon has potential applications in the separation\cite{sajeesh2014particle}, trapping, and filtration of particles, presenting a novel approach to particle manipulation under surface-driven flow through nano-tubes across a spectrum of conditions.

The fluid flow through nano-channels happens in a very low Reynolds number (Re) regime. This opens up the possibility of developing a systematic perturbation approach for the surface-driven effects on fluid flow through such channels, where Re could be used as the small parameter for the perturbation scheme. We use this scheme to understand the effects of surface undulations on flow through soft cylindrical boundaries. The zeroth order flow is considered Poiseuille. In the first order, considering a traveling Fourier mode of surface undulation as the source term, correction to the zeroth order flow is found. This marks a cross-over in the radial directions between the Poiseuille flow and the flow induced by surface undulations. The cross-over could be controlled by the surface drive and it particularly depends on the nature of the fluid. 

\par The effect of non-linearity and dynamics of the Navier-Stokes equation shows up in the second order in perturbation. We extend the analysis to this order to obtain closed-form information on the feedback of the flow induced on the soft boundary. The existence of the flow in the first order in perturbations acting as a source in the second order induces finite secondary excitations to the soft boundary due to the nonlinearity of the dynamics. This perturbative analysis, being essentially linear, is done to explore the effects of a single source mode of surface undulations. However, this is enough in the linear regime to represent the effects of all Fourier modes of surface undulations that could be there if the surface is driven by any general drive.

We organize the paper as follows: first, we present the Navier-Stokes equation\cite{kambe2007elementary, landau1959fluid} governing fluid flow and the associated boundary conditions for the geometry relevant to our model system. Subsequently, we solve the Navier-Stokes equation under the low Reynolds number approximation, a justification for which is provided in the subsequent discussion. We identify the regime where the first-order correction becomes comparable to the zeroth-order flow, describing the transition from pressure-driven flow to boundary-driven flow. In the final sections, we present conclusive remarks and discussions about the implications and potential applications of our results. 

\section{NAVIER-STOKES EQUATION}
Our interest in this paper is to have a detailed understanding of the structure of fluid flow through a cylindrical geometry under the joint existence of pressure gradient along the axis of the tube and surface fluctuations and identify the dominant flow regimes along the radius of the cylindrical surface. To that goal, we intend to get solutions of the Navier-Stokes equation of such a system on nanometer scales (for nano-tube). We will identify in this section the fact that such flows through nano-tubes, even when surface fluctuations are of frequency of several thousands of Hertz, are in the low Reynolds number regime. Based on this fact, we develop a linear perturbation analysis of the flow. The linearity of the theory allows us to work with a single mode of surface fluctuations where, under the general condition of the existence of several Fourier modes, superposition works. 

Navier-Stokes equation for the velocity profile $\mathbf{u}(\mathbf{r}, t)$ of a fluid flow in its most general form is
\begin{multline}
    \rho \left(\frac{\partial {\bf u}}{\partial t}  + (\bf {u}.\nabla)\bf {u}\right) = -\nabla P  + \eta \nabla^2 \bf{u}+\\ \left(\frac{1}{3}\eta + \zeta\right)\nabla(\nabla . {\bf u}) + {\bf \emph f_{ext}},\label{eq:1}
\end{multline}
supplemented with continuity equation:
\begin{equation}
    \frac{\partial \rho}{\partial t} + \nabla .(\rho \mathbf{u}) = 0,\label{eq:2}
\end{equation}
where $\rho$ and $\eta$ are respectively, mass density and dynamic viscosity of the fluid, and $\zeta$ is known as the second viscosity coefficient or bulk viscosity of the fluid. In eq.(\ref{eq:1}), $\bf\emph f_{ext}$ is the body force acting on the fluid. 

The model system that we are going to consider is fluid flow in a cylindrical nano-channel of average radius $R_0$ whose walls are oscillating in space and time about its average radius $R_0$.  Here we will examine the effects of sinusoidal forcing characterized by specific frequency $\omega$ and wavenumber $k$ mainly in traveling wave configuration which can induce dragging of particles along a traveling wave front in the fluid. 

Throughout this paper, we will assume the in-compressibility of fluid, (i.e., $\frac{d\rho}{dt} = 0$) which gives $\nabla. {\bf{u}} = 0 $ due to the continuity equation. Due to this assumption third term in the Navier-Stokes equation drops out due to its dependence on the $\nabla \cdot \mathbf{u}$ term. Thus above equations reduce to the following form:
\begin{equation}
    \rho \left(\frac{\partial \bf{u}}{\partial t}  + (\bf{u}.\nabla)\bf{u}\right) = -\nabla P  + \eta \nabla^2 \bf{u}+ {\bf \emph f_{ext}}\label{eq:3}
\end{equation}
\par and
\begin{equation}
    \nabla .{\bf{u}} = 0.\label{eq:4}
\end{equation}

\par Since we intend to solve for the velocity profile $\bf{u}$ by perturbation theory, to make a comparison of different terms of the Navier-Stokes equation, we will non-dimensionalise the equation. Introducing $r' = r/R_0$, $t' = t/\tau$ ($\tau$ = $2\pi/\omega$), ${\bf{u'}} = {\bf{u}}/U (U = R_0/\tau = \omega R_0/2\pi)$, $P' = P/P_0$ ($P_0 = \eta U/R_0$) as dimensionless variables. With dimensionless quantities, the eq.(\ref{eq:3}) becomes:
\begin{equation}
    Re \left( \frac{\partial \bf{u'}}{\partial t'} + (\bf{u'}.\nabla')\bf{u'}  \right) = -\nabla' P' +  \nabla'^2{\bf u'} + \; \frac{2\pi R_0}{\eta \omega}{\bf \emph f_{ext}}\label{eq:5},
\end{equation}
where $Re = \rho U R_0/\eta = \rho \omega R_0^2/2\pi\eta$ is Reynolds number. Typical numerical values considered in our analysis are $\rho \sim 10^3\;  \text{kg/m}^3$, $\omega \sim 1000\; \text{Hz}, R_0 \sim 100 \; \text{nm}$, and $\eta \sim 10^{-3} \; \text{Pa s}$. This results in a calculated Reynolds number of approximately $Re \sim 10^{-6}$, which is an extremely small value.

\par Since we are working in a low Reynolds number regime, $Re$ will serve as a perturbation parameter. Thus, replacing $Re$ by $\epsilon$ eq.(\ref{eq:5}) takes the shape:
\begin{equation}
     \epsilon \left( \frac{\partial \bf{u'}}{\partial t'} + (\bf{u'}.\nabla')\bf{u'}  \right) = -\nabla' P' +  \nabla'^2{\bf u'} + \; \frac{2\pi R_0}{\eta \omega}{\bf \emph f_{ext}}\label{eq:6}.
\end{equation}
Moreover, since the geometry of our system is cylindrical with an undulating surface, it is convenient to use cylindrical coordinates. In cylindrical coordinates ${\bf{u}} = v \boldsymbol{\hat{r}} + u \boldsymbol{\hat{z}}$ (assuming axis-symmetry). 

Now, substituting the expression for Laplacian and convective derivatives in cylindrical coordinates in the above equation and separating the vector eq.(\ref{eq:6}) into its component equations we get two equations. The equation for $\boldsymbol{\hat{z}}$-component is 
\begin{multline}
    \epsilon\left( \frac{\partial u'}{\partial t'} + u' \frac{\partial u'}{\partial z'} + v' \frac{\partial u'}{\partial r'}\right) = -\frac{\partial P'}{\partial z'} +  \frac{\partial^2u'}{\partial z'^2} + \frac{\partial^2 u'}{\partial r'^2} \\+ \frac{1}{r'}\frac{\partial u'}{\partial r'} + \frac{2\pi R_0}{\eta \omega} \;f_z\label{eq:7}
\end{multline}
and $\boldsymbol{\hat{r}}$ equation is
\begin{multline}
    \epsilon \left( \frac{\partial v'}{\partial t'} + v' \frac{\partial v'}{\partial r'} + u' \frac{\partial v'}{\partial z'}\right) =  \frac{\partial^2v'}{\partial z'^2} + \frac{\partial^2 v'}{\partial r'^2} \\+ \frac{1}{r'}\frac{\partial v'}{\partial r'} - \frac{v'}{r'^2} + \frac{2\pi R_0}{\eta \omega} \;f_r.\label{eq:8}
\end{multline}
\par Where we have used ${\bf \emph f_{ext}} = f_r \boldsymbol{\hat{r}} + f_z \boldsymbol{\hat{z}}$. Expanding $u'$ and $v'$ in perturbation series:

\begin{equation}
    u' = u'_0 + \epsilon u'_1 +\epsilon^2 u'_2 + \dots , \label{eq:9}
\end{equation}

\begin{equation}
     v' = v'_0 + \epsilon v'_1 +\epsilon^2 v'_2 +\dots , \label{eq:10}
\end{equation}
and Similarly, the radius of the nano-channel can be written as:
\begin{equation}
    R(z,t) = R_0 + \epsilon R_1(z,t) + \epsilon^2 R_2(z,t)+\dots , \label{eq:11}
\end{equation}
where $R_0$ is the average radius of the nano-channel and $R_1$, and $R_2$ are respectively first- and second-order undulations present on the surface of the nano-channel, for example.

\subsection{Boundary conditions}
To determine the solution uniquely one needs to impose appropriate boundary conditions on the velocity profile. Boundary conditions describe the behavior of fluid at boundaries that the solution has to obey at all orders. The velocity profile $\mathbf{u}(\mathbf{r}, t)$ of the fluid satisfies the following boundary condition (known as the Kinematic boundary condition):
\begin{multline}
    v(r = R(z,t), z,  t) -  u(r = R(z,t), z, t) \frac{\partial R(z,t)}{\partial z} - \\\frac{\partial R(z,t)}{\partial t} = 0. \label{eq:12}
\end{multline}

\subsection{Zeroth order equations}
\par We assume that forcing ($f_r$ and $f_z$) which is induced by boundary fluctuations that is orders of magnitude smaller than the pressure gradient, is present at higher orders only. The leading order transport of fluid is due to the pressure gradient. Using eq.(\ref{eq:7}) and (\ref{eq:8}) one can write out the zeroth order equations:
\begin{equation}
     -\frac{\partial P'}{\partial z'} + \frac{\partial^2u'_0}{\partial z'^2} + \frac{\partial^2 u'_0}{\partial r'^2} + \frac{1}{r'}\frac{\partial u'_0}{\partial r'} = 0, \label{eq:13}
\end{equation}
and 
\begin{equation}
     \frac{\partial^2v'_0}{\partial z'^2} + \frac{\partial^2 v'_0}{\partial r'^2} + \frac{1}{r'}\frac{\partial v'_0}{\partial r'} - \frac{v'_0}{r'^2} = 0.\label{eq:14}
\end{equation}

\par Now zeroth-order equations solve for steady flow along $z$ - direction in a much simple setting
\begin{equation}
     \frac{\partial u'_0}{\partial z'} =0\;,\;\frac{\partial^2 u'_0}{\partial z'^2} = 0\; \text{and} \; v'_0 = 0.\label{eq:15}
\end{equation}
 Thus equation for $u'_0$ becomes: 
 \begin{equation}
     -\frac{\partial P'}{\partial z'} + \frac{\partial^2 u'_0}{\partial r'^2} + \frac{1}{r'}\frac{\partial u'_0}{\partial r'} = 0.\label{eq:16}
\end{equation}
Assuming a constant pressure gradient, solving the equation above yields:
\begin{equation}
    u'_0 = C_1 + \frac{1}{4}\frac{\partial P'}{\partial z'}r'^2, \label{eq:17}
\end{equation}
where $C_1$ is the integration constant. Restoring the dimensions of various quantities, we have:
\begin{align}
    u_0 &= U u'_0 \notag\\ & = U \left( C_1 + \frac{1}{4\eta}\frac{\partial P}{\partial z}\frac{R_0}{P_0}r^2\frac{1}{R_0^2} \right) \notag\\
    &= C_2 + \frac{1}{4\eta}\frac{\partial P}{\partial z} r^2. \label{eq:18}
\end{align}

Since $v'_0 = 0$, 
\begin{equation}
    v_0 = 0. \label{eq:19}
\end{equation}
Now, applying the no-slip boundary condition to eq.(\ref{eq:18}), which is consistent with the kinematic boundary condition in general in the absence of surface undulations, we have $u_0 = 0$ at $r = R_0$ to give
\begin{equation}
    u_0 = -\frac{1}{4\eta}\frac{\partial P}{\partial z}\left(R_0^2 - r^2\right),\label{eq:20}
\end{equation}
which is the standard Poisuille's flow profile under a constant pressure gradient. Note that the above zeroth order solution is valid for any Reynolds number as long as condition in eq.(\ref{eq:15}) are met.

The continuity equation at zeroth order is
\begin{equation}
     \frac{1}{r}\frac{\partial}{\partial r}(rv_0) + \frac{\partial u_0} {\partial z} = 0.\label{eq:21}
\end{equation}
With these expressions for $u_0$ and $v_0$, the zeroth-order continuity equation trivially satisfies, as the pressure gradient $\partial P/\partial z $ is a constant.

\subsection{First order equations}
First order equations obtained from eq.(\ref{eq:7}) and (\ref{eq:8}) are:
\begin{equation}
     \frac{\partial^2 u'_1}{\partial z'^2} + \frac{\partial^2 u'_1}{\partial r'^2} + \frac{1}{r'}\frac{\partial u'_1}{\partial r'} = -\frac{2 \pi R_0}{\eta \omega}f_z^{(1)},\label{eq:22}
\end{equation}
and
\begin{equation}
     \frac{\partial^2 v'_1}{\partial z'^2} + \frac{\partial^2 v'_1}{\partial r'^2} + \frac{1}{r'}\frac{\partial v'_1}{\partial r'} - \frac{v'_1}{r'^2} = -\frac{2 \pi R_0}{\eta \omega}f_r^{(1)},\label{eq:23}
\end{equation}
where $f_z^{(1)}$ and $f_r^{(1)}$ respectively represent the $z$ and $r$ components of forces experienced by the fluid due to the surface undulations present on the walls of nano-channels. We consider here these forces to be of traveling wave form due to the existence of traveling wave modes on the surface of the fluid which would get automatically justified in what follows.

Let us take $f_r = \epsilon f_r^{(1)} + \epsilon^2 f_r^{(2)} =(\epsilon f_1(r) + \epsilon^2 f_2(r))\cos(\omega t-kz)$, where $f_r^{(1)} = f_1(r)\cos(\omega t -kz)$ and $f_r^{(2)} = f_2(r)\cos(\omega t - kz)$. Here $f_1(r)$ and $f_2(r)$ are $r$ dependent parts of forcing yet to be identified. Thus, eq.(\ref{eq:23}) becomes:
\begin{align}
    \frac{\partial^2 v'_1}{\partial z'^2} &+ \frac{\partial^2 v'_1}{\partial r'^2} + \frac{1}{r'}\frac{\partial v'_1}{\partial r'} - \frac{v'_1}{r'^2} \nonumber \\
    &= -\frac{2\pi R_0}{\eta \omega} f_1(r'R_0)\cos(2\pi t'-k' z'),  \label{eq:24}
\end{align}

where $k' = k R_0$. It is clear from above eq.(\ref{eq:24}) that solution for $v'_1$ will be of the form $v'_1 = g(r')\cos(2\pi t' - k'z')$. Where $g(r')$ is to be determined. Substituting this ansatz for $v'_1$ in eq.(\ref{eq:24}) we get an equation for $g(r')$. 
\begin{multline}
    \frac{d^2g(r')}{dr'^2} + \frac{1}{r'}\frac{dg(r')}{dr'} - \left(k^2R_0^2 + \frac{1}{r'^2}\right)g(r') \\= -\frac{2\pi R_0}{\eta \omega}f_1(r' R_0)\label{eq:25}
\end{multline}
\par We take $f_1(r'R_0) = 0$, thus above equation becomes:
\begin{equation}
    \frac{d^2g(r')}{dr'^2} + \frac{1}{r'}\frac{dg(r')}{dr'} - \left(k^2R_0^2 + \frac{1}{r'^2}\right)g(r') = 0.\label{eq:26}
\end{equation}
Which is modifies Bessel's equation. Therefore its solution is:
\begin{equation}
    g(r') = c_1 I_1(k R_0 r') + c_2 K_1(k R_0 r') .\label{eq:27}   
\end{equation}

\par Where $I_1(k R_0 r')$ and $K_1(k R_0 r')$ are modified Bessel functions of first and second kinds, respectively. Since in the above equations, $r'$ represents the radial distance in nano-channel, thus we can use the approximation of $r'$ being small to simplify our expressions. Using the approximations of $I_1$ and $K_1$ for small arguments:

\begin{equation}
     I_1(k R_0 r') \sim \frac{k R_0 r'}{2},\label{eq:28}
\end{equation}
\begin{equation}
     K_1(k R_0 r') \sim \frac{1}{k R_0 r'}\label{eq:29}
\end{equation}

eq.(\ref{eq:27}) simplifies to
\begin{equation}
    g(r') = c_1 \biggl(\frac{kR_0r'}{2}\biggl)  + c_2 \biggl(\frac{1}{k R_0 r'}\biggl) .\label{eq:30}   
\end{equation}

\par Since the second term diverges at $r'=0$, we set $c_2 = 0$. Thus, we get:

\begin{equation}
    g(r') = c_1 \biggl(\frac{kR_0r'}{2}\biggl).\label{eq:31}   
\end{equation}
\par Therefore
\begin{equation}
     v'_1 = c_1 \biggl(\frac{kR_0r'}{2}\biggl)\cos(2\pi t' - k' z').\label{eq:32}
\end{equation}
\par Now using the continuity equation in first order we find out $u'_1$ from the expression of $v'_1$.

\begin{equation}
    u'_1 = c_1\biggl(\sin(2 \pi t'-k'z')-\sin(2\pi t')\biggl) .\label{eq:33}  
\end{equation}

\par Since $f_z = \epsilon f_z^{(1)} + \epsilon^2 f_z^{(2)}$. Therefore
\begin{equation}
     \frac{\partial^2 u'_1}{\partial z'^2} + \frac{\partial^2 u'_1}{\partial r'^2} + \frac{1}{r'}\frac{\partial u'_1}{\partial r'} = -\frac{2\pi R_0}{\eta \omega}f_z^{(1)}.\label{eq:34}
\end{equation}
Which gives,
\begin{equation}
    f_z^{(1)} = c_1 \frac{\eta \omega k'^2}{2 \pi R_0}\sin(2\pi t'-k' z').\label{eq:35}
\end{equation}

\par Now after restoring the dimensions of various terms we get

\begin{equation}
    u_1(r,z,t) = c_1 \frac{\omega R_0}{2 \pi}\biggl(\sin(\omega t-k z)-\sin(\omega t)\biggl) .\label{eq:36}  
\end{equation}
\begin{equation}
     v_1 = c_1 \biggl(\frac{k \omega R_0}{4 \pi}\biggl) r \cos(\omega t - k z).\label{eq:37}
\end{equation}

and

\begin{equation}
    f_z^{(1)} = \frac{\eta \omega k^2 R_0}{2 \pi}\sin(\omega t-kz) \label{eq:38}
\end{equation}

Finally using the Kinematic boundary condition to find $R_1(z,t)$ we have,

\begin{equation}
    R_1(z,t) = \frac{c_1 k R_0^2}{4 \pi}\sin(\omega t-kz)\label{eq:39}.
\end{equation}

\par Calling the amplitude of the surface undulations $\displaystyle{\frac{c_1kR_0^2}{4\pi}}$ as $A_0$, we have:

\begin{equation}
    R_1(z,t) =  A_0 \sin(\omega t - kz).\label{eq:40}
\end{equation}

Thus,

\begin{equation}
    c_1 = \frac{4\pi A_0}{kR_0^2}.\label{eq:41}
\end{equation}

In terms of $A_0$ we get,

\begin{equation}
    u_1(r,z,t) = \frac{2\omega A_0
    }{kR_0}\biggl(\sin(\omega t-k z)-\sin(\omega t)\biggl)\label{eq:42}
\end{equation}
and
\begin{equation}
    v_1(r,z,t) = \frac{\omega A_0}{R_0} r \cos(\omega t - kz).\label{eq:43}
\end{equation}

\par From above we can calculate the expressions of the ratio of amplitudes of $u_1$ and $v_1$.

\begin{equation}
    \Biggl|\dfrac{u_1}{v_1}\Bigg|\sim \dfrac{4\dfrac{\omega}{k}\dfrac{A_0}{R_0}}{\dfrac{\omega A_0}{R_0}r}.\label{eq:44} 
\end{equation}

or,

\begin{equation}
    \Biggl|\dfrac{u_1}{v_1}\Bigg|\sim \frac{1}{kr} \sim \frac{\lambda}{r}.\label{eq:45}
\end{equation}

\par Thus by simply increasing or decreasing the wavelength of surface undulations one can simply make one component dominate over the other. 

\section{CROSSOVER OF SURFACE INDUCED FLOW AND BULK FLOW}
\par To find out the value of $r$ where first order correction $u_1$ becomes comparable to base flow $u_0$, we set,

\begin{equation}
    |u_0|\bigg|_{r=r_0} \approx Re |u_1|\bigg|_{r=r_0}.\label{eq:46}
\end{equation}
Which gives,

\begin{equation}
    \frac{1}{4\eta}\left|\frac{\partial P}{\partial z}\right| (R_0^2 - r_0^2) \approx 4 Re\frac{\omega}{k}\frac{A_0}{R_0} \label{eq:47}     
\end{equation}

or

\begin{equation}
    \frac{1}{4 \eta}\left|\frac{\partial P}{\partial z}\right| \biggl(1-\frac{r_0^2}{R_0^2}\biggl)R_0^2 \approx 4Re \frac{\omega}{k}\frac{A_0}{R_0} .\label{eq:48}
\end{equation}
Solving gives :
\begin{equation}
    \frac{r_0^2}{R_0^2} \approx 1-Re \frac{16 \omega A_0 \eta}{k R_0^3}\left|\frac{\partial P}{\partial z}\right|^{-1}.\label{eq:49}
\end{equation}

Since, $Re = \rho \omega R_0^2/2\pi \eta$. Therefore

\begin{equation}
    \frac{r_0^2}{R_0^2} \approx 1-\frac{4 \rho \omega^2 A_0 \lambda}{\pi^2 R_0 }\left|\frac{\partial P}{\partial z}\right|^{-1}.\label{eq:50}
\end{equation}
Giving,
\begin{equation}
    \frac{r_0}{R_0} \approx \sqrt{1-\frac{4 \rho \omega^2 A_0 \lambda}{\pi^2 R_0}\left|\frac{\partial P}{\partial z}\right|^{-1}}.\label{eq:51}
\end{equation}
We define $Q_0$ as the dimensionless ratio $R_0/r_0$, representing the extent to which the fluid flow is boundary-driven. A higher value of the $R_0/r_0$ ratio signifies a boundary-dominant flow, while a smaller value of $Q_0$ indicates a pressure-driven flow.

\begin{equation}
    Q_0 = \frac{R_0}{r_0} .\label{eq:52}
\end{equation}
\begin{figure}[ht]
% For example, with the graphics package use
\includegraphics[width=8cm,height=7cm]{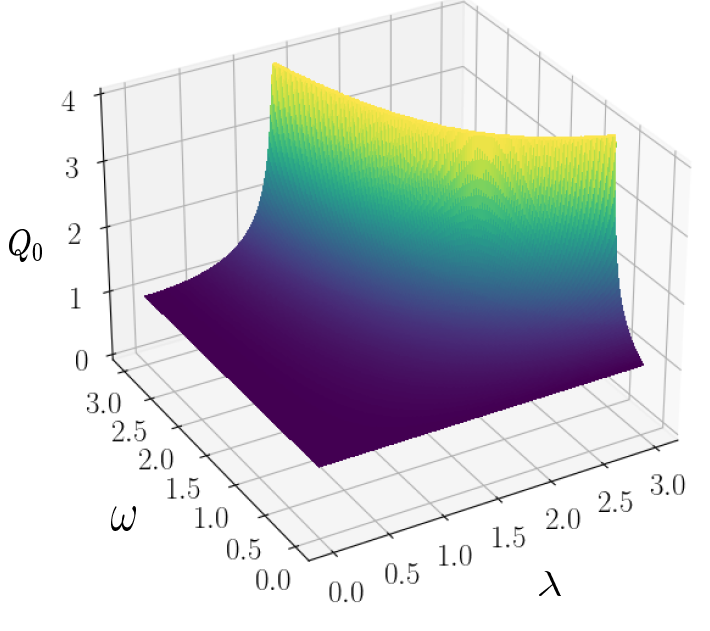}
\caption{Variation of $Q_0$ with $\omega$(in 10$^{5}$ rad s$^{-1}$) and $\lambda$(in 10$^{-7}$ m)($R_0 = 100$ nm, $A_0 = 10 $ nm, $|\partial P/\partial z| = 10^5$ Pa/m and $\rho$ = 0.5 $\times$ 10$^3$ kg/m$^3$)}
\label{fig:1}       % Give a unique label
\end{figure}

\begin{figure}[ht]
% For example, with the graphics package use
\includegraphics[width=8cm,height=7cm]{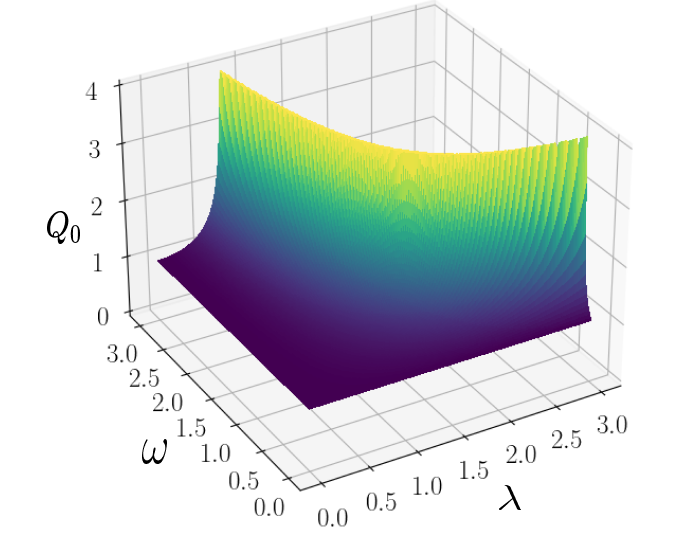}
\caption{Variation of $Q_0$ with $\omega$(in 10$^{5}$ rad s$^{-1}$) and $\lambda$(in 10$^{-7}$ m)($R_0 = 100$ nm, $A_0 = 10 $ nm, $|\partial P/\partial z| = 10^5$ Pa/m and $\rho$ = 1.5 $\times$ 10$^3$ kg/m$^3$)}
\label{fig:2}       % Give a unique label
\end{figure}

\par In the fig.(\ref{fig:1}) and (\ref{fig:2}) shown above, we have presented a surface plot of the dimensionless parameter $Q_0$ against driving parameters $\omega$ and $\lambda$ for two different values of fluid density. For this, we have taken the typical values of the other parameters, i.e., $R_0 = 100$ nm, $A_0 = 10$ nm, $|\partial P/\partial z| = 10^5 $ Pa/m. This plot essentially specifies the values of driving parameters $\omega$ and $\lambda$ needed for a high value of $Q_0$, i.e., the fluid flow will be highly boundary-driven rather than pressure-driven.

\vspace{\baselineskip}

\section{Second order equations}
\par Before concluding the paper, we would like to digress a little by presenting the second-order equations and their solutions as the effects of the time dependence of the Navier-Stokes equation will show up in second-order under this perturbation scheme. Boundary conditions considered here take into account a soft boundary. The flow will generate, in the second order, feedback to the boundary undulations considered in the previous order. The soft boundary admitting those feedback modes would keep the dynamics closed. Using eq.(\ref{eq:7}) and (\ref{eq:8}) we obtain the second-order equations. Here, the right side is completely known in terms of previously calculated quantities and their derivatives.
\begin{multline}
      \frac{\partial ^2u'_2}{\partial z'^2}  +\frac{\partial^2u'_2}{\partial r'^2}+ \frac{1}{r'}\frac{\partial u'_2}{\partial r'} + \frac{2 \pi R_0}{\eta \omega}f_z^{(2)}\\ = \frac{\partial u'_1}{\partial t'} + u'_0 \frac{\partial u'_1}{\partial z'} + v'_1 \frac{\partial u'_0}{\partial r'}.\label{eq:53}
\end{multline}

\begin{multline}
     \frac{\partial ^2v'_2}{\partial z'^2}+\frac{\partial^2v'_2}{\partial r'^2}+ \frac{1}{r'}\frac{\partial v'_2}{\partial r'} - \frac{v'_2}{r'^2} + \frac{2\pi R_0}{\eta \omega} f_r^{(2)} \\  = \frac{\partial v'_1}{\partial t'} + u'_0 \frac{\partial v'_1}{\partial z'}.\label{eq:54}
\end{multline}

\par We take  $f_z^{(2)}$ and $f_r^{(2)} $ to be the forces that have to be identified self-consistently. We take $\displaystyle f_r^{(2)} = k'^2 \frac{\eta \omega}{R_0} \left(\frac{A_0}{R_0}\right)^2 \sin(k'z')$. The reason for choosing this particular form of $f_r^{(2)}$ is to avoid the divergence that will appear in the structure of $R_2$ otherwise. Thus eq.(\ref{eq:54}) becomes:

\begin{multline}
     \frac{\partial ^2v'_2}{\partial z'^2}+\frac{\partial^2v'_2}{\partial r'^2}+ \frac{1}{r'}\frac{\partial v'_2}{\partial r'} - \frac{v'_2}{r'^2} =   -2\pi k'^2 \frac{A_0^2}{R_0^2} \sin(k'z') \\ +\frac{\partial v'_1}{\partial t'} + u'_0 \frac{\partial v'_1}{\partial z'}.\label{eq:55}
\end{multline}
\par Where $r' = r/R_0, t' = t/\tau \;\text{and} \;k' = kR_0$ as before. Substituting the expressions for $u'_0 \;\text{and} \;v'_1$, above equation takes the form:

\begin{multline}
     \frac{\partial^2 v'_2}{\partial z'^2} + \frac{\partial^2 v'_2}{\partial r'^2} + \frac{1}{r'}\frac{\partial v'_2}{\partial r'} - \frac{v'_2}{r'^2} = -2\pi k'^2 \frac{A_0^2}{R_0^2} \sin(k'z') \\- c_1k' \sin(2\pi t' - k'z') \left(\pi + \frac{1}{8}\frac{\partial P'}{\partial z'}k'\right)r' \\ + \frac{1} {8}\frac{\partial P'}{\partial z'}c_1 k'^2 \sin(2\pi t' - k'z')r'^3.\label{eq:56}
\end{multline}
Which can also be written as
\begin{multline}
     \frac{\partial^2 v'_2}{\partial z'^2} + \frac{\partial^2 v'_2}{\partial r'^2} + \frac{1}{r'}\frac{\partial v'_2}{\partial r'} - \frac{v'_2}{r'^2} = -2\pi k'^2 \frac{A_0^2}{R_0^2} \sin(k'z') \\+ (B r' + C r'^3) \sin(2\pi t' - k' z').\label{eq:57}
\end{multline}
With $B = -c_1 k'\displaystyle\left(\pi + \frac{1}{8}\frac{\partial P'}{\partial z'} k'\right) \;\text{and}\; C = \displaystyle\frac{1}{8}\displaystyle\frac{\partial P'}{\partial z'}c_1k'^2$.

Assuming a solution of the form $v'_2 = h(r') \sin(2\pi t' - k' z' ) + 2\pi (A_0^2/R_0^2) \sin(k'z')$ and substituting in the above equation we get the following equations for $h(r')$:

\begin{multline}
    \frac{d^2h(r')}{d r'^2} + \frac{1}{r'}\frac{dh(r')}{dr'} - \biggl(k'^2 + \frac{1}{r'^2}\biggl)h(r')  = Br'+Cr'^3.\label{eq:58}
\end{multline}
The above equation can be solved using the method of variation of parameters and the solution is $\displaystyle h(r') = \frac{B}{8}r'^3 + \frac{C}{24}r'^5$.
Thus,

\begin{multline}
    v'_2 =  \biggl(\frac{B}{8}r'^3 + \frac{C}{24}r'^5\biggl) \sin(2 \pi t' - k' z') + 2\pi \frac{A_0^2}{R_0^2} \sin(k'z').\label{eq:59}
\end{multline}

\par Restoring the dimensions gives:
\begin{multline}
    v_2(r,z,t) = \frac{\omega R_0}{2 \pi}\left(\frac{B}{8}\left(\frac{r}{R_0}\right)^3 + \frac{C}{24}\left(\frac{r}{R_0}\right)^5\right)\\ \times\sin(\omega t - kz) + \frac{\omega A_0^2}{R_0} \sin(kz).\label{eq:60}
\end{multline}

To find out $u_2$ we use the continuity equation in the second order. The continuity equation in second order is:
\begin{equation}
    \frac{1}{r}\frac{\partial}{\partial r}(rv_2) + \frac{\partial u_2}{\partial z} = 0.\label{eq:61}    
\end{equation}

Which gives:

\begin{multline}
u_2(r,z,t) = \frac{\omega R_0}{2 \pi}
 \biggl(\frac{B}{2kR_0}\left(\frac{r}{R_0}\right)^2 + \frac{C}{4kR_0}\left(\frac{r}{R_0}\right)^4\biggl) \\ \times \left(\cos(\omega t - k z) - \cos(\omega t)\right).\label{eq:62}
\end{multline}

\par Where $\displaystyle B = -c_1 k R_0 \left(\pi + \frac{\pi k R_0^2}{4\eta \omega} \frac{\partial P}{\partial z}\right)$ and $ \displaystyle C = c_1 \frac{\pi k^2 R_0^3}{4\eta \omega} \frac{\partial P}{\partial z}$.

\par Substituting the value of $c_1$ in terms of $A_0$ we obtains, $\displaystyle B = -\frac{4\pi^2 A_0}{R_0}\left(1 + \frac{               k R_0^2}{4\eta \omega} \frac{\partial P}{\partial z}\right)$ and $\displaystyle C = \frac{k \pi A_0 R_0}{4 \eta\omega}\frac{\partial P}{\partial z}$.

\par To find out the $R_2(z,t)$ we use Kinematic boundary condition in second order.

\begin{equation}
    v_2 - u_1 \frac{\partial R_1}{\partial z} - u_0 \frac{\partial R_2}{\partial z} - u_2 \frac{\partial R_0}{\partial z} -\frac{\partial R_2}{\partial t}\bigg|_{r=R_0} = 0.\label{eq:63}
\end{equation}

Since $u_0 = 0$ at $r = R_0$ and $R_0$ is constant, thus
\begin{equation}
    v_2 - u_1 \frac{\partial R_1}{\partial z} -\frac{\partial R_2}{\partial t}\bigg|_{r=R_0} = 0.\label{eq:64}
\end{equation}

Substituting various expressions at $r = R_0$, we obtains:

\begin{multline}
    R_2(z,t) = -\frac{R_0}{2\pi}\left(\frac{B}{8} + \frac{C}{24}\right)\cos(\omega t - kz) \\- \frac{A_0^2}{R_0}\sin(kz/2) \sin(2 \omega t - 3kz/2).\label{eq:65}
\end{multline}

Thus we have:
\begin{multline}
    R(z,t) = R_0 + Re R_1(z,t) + Re^2 R_2(z,t) \\ = R_0 + Re A_0 \sin(\omega t - kz) -\\Re^2 \frac{R_0}{2\pi}\biggl(\frac{B}{8}+\frac{C}{24}\biggl)\cos(\omega t-kz) \\- Re^2\frac{A_0^2}{R_0}\sin(kz/2)\cos(2\omega t - 3 kz/2).\label{eq:66}
\end{multline}

\begin{multline}
    R(z,t) = R_0 + Re R_1(z,t) + Re^2 R_2(z,t) \\ = R_0 + Re A_0 \sin(\omega t - kz) + Re^2 \frac{R_0}{16}\\ \times\biggl(4\pi\frac{A_0}{R_0}+\left(\pi - \frac{1}{12}
    \right)\frac{k A_0 R_0}{\eta \omega}\frac{\partial P}{\partial z}\biggl)\cos(\omega t-kz) \\- Re^2\frac{A_0^2}{R_0}\sin(kz/2)\cos(2\omega t - 3 kz/2).\label{eq:67}
\end{multline}

In this structure of the surface undulations, the order $Re$ term could be imagined to be the driving of the surface by some external agent where the free parameter is $c_1$ which is proportional to the driving amplitude $A_0$. At the order $Re^2$, the first term is the feedback on the soft surface due to the dynamics whereas the second term is that due to the non-linearity of the N-S equation. Putting in place the expression for $Re$ and $c_1$ one can now get the $R(z,t)$ in terms of all physical parameters as:

\begin{multline}
    R(z,t) = R_0 + Re R_1(z,t) + Re^2 R_2(z,t) \\ = R_0 + \frac{\rho \omega R_0^2}{2 \pi \eta} A_0 \sin(\omega t - kz) + \frac{\rho^2 \omega^2 R_0^5}{64 \pi^2 \eta^2}\\ \times\biggl(4\pi\frac{A_0}{R_0}+\left(\pi - \frac{1}{12}
    \right)\frac{k A_0 R_0}{\eta \omega}\frac{\partial P}{\partial z}\biggl)\cos(\omega t-kz) \\- \frac{\rho^2 \omega^2 R_0^3 A_0^2}{4 \pi^2 \eta^2}\sin(kz/2)\cos(2\omega t - 3 kz/2).\label{eq:68}
\end{multline}

\section{DISCUSSION}

\par

\par In this paper, within the framework of perturbation theory, we have explored soft-surface driven transport in an undulating nano-channel along with pressure-driven transport. Employing the advantages of a low Reynolds number regime, we provide a detailed exploration of boundary-driven transport. This mechanism is significant for applications in particle separation and filtration processes. We have shown where the surface-driven flow surpasses pressure-driven flow, and we have quantified the surface-driven or pressure-driven nature of flow in the undulating nano-channel using the parameter $Q_0$ and shown a surface plot of $Q_0$ in terms of forcing parameters $\omega$ and $\lambda$.

\par The quantification by $Q_0$ is crucial from an engineering perspective. This single parameter takes into account the fluid present in the nano-channel i.e., the density of the fluid present as well as the frequency and wavelength of the forcing applied. This is essential from an engineering perspective as it allows us to determine which type of forcing is needed for what kind of fluid. These dependencies are essential in optimising the design and function of such nano-scale systems.

We have shown that by simply increasing or decreasing the wavelength of driving one can make the longitudinal component of transport dominate over transverse. Such modes of fluid could be tuned to drag particles in the fluid against a pressure-driven flow to engineer efficient filtration.

\section{ACKNOWLEDGEMENTS}
Aakash Anand would like to thank the Council of Scientific and Industrial Research (CSIR), India for providing funding during this research.
\bibliography{references.bib}

\end{document}